\documentclass[sigconf,screen]{acmart}
\AtBeginDocument{%
  }

\setcopyright{acmlicensed}
\copyrightyear{2018}
\acmYear{2018}
\acmDOI{XXXXXXX.XXXXXXX}
\acmConference[Conference acronym 'XX]{Make sure to enter the correct
  conference title from your rights confirmation email}{June 03--05,
  2018}{Woodstock, NY}
\acmISBN{978-1-4503-XXXX-X/2018/06}

\usepackage{amssymb}
\usepackage{enumitem}

\usepackage{array}
\usepackage{colortbl}
\usepackage{xcolor}
\usepackage{pifont}
\usepackage{makecell}
\usepackage{adjustbox}
\usepackage{tikz}
\usepackage{fontawesome5}
\usepackage{wasysym} 
\usepackage{graphicx}
\usepackage{tabularx}
\usepackage{url}

\definecolor{darkgreen}{RGB}{0,128,0}  
\definecolor{darkred}{RGB}{204,0,0}    

\usepackage[utf8]{inputenc}
\usepackage{caption}
\usepackage[most]{tcolorbox}
\tcbuselibrary{listingsutf8}

\tcbset{
  mybox/.style={
    colback=#1,
    colframe=white,
    boxrule=0pt,
    frame hidden,
    interior hidden,
    sharp corners, 
    left=6pt,  
    right=6pt  
  },
  mytitle/.style={
    colback=#1,
    colframe=black,
    boxrule=0.4pt,
    fontupper=\bfseries,
    coltext=black,
    boxsep=2pt,
    left=1pt, right=1pt, top=2pt, bottom=2pt,
    sharp corners
  }
}

\hypersetup{
    colorlinks=true,
    linkcolor=blue,
    citecolor=blue,
    urlcolor=blue
}

\copyrightyear{2026}
\acmYear{2026}
\setcopyright{cc}
\setcctype{by}
\acmConference[CAIN '26]{2026 IEEE/ACM 5th International Conference on AI Engineering - Software Engineering for AI}{April 12--13, 2026}{Rio de Janeiro, Brazil}
\acmBooktitle{2026 IEEE/ACM 5th International Conference on AI Engineering - Software Engineering for AI (CAIN '26), April 12--13, 2026, Rio de Janeiro, Brazil}
\acmDOI{10.1145/3793653.3793801}
\acmISBN{979-8-4007-2475-6/2026/04}

\begin{document}

\title{Toward Architecture-Aware Evaluation Metrics for LLM Agents}

\settopmatter{authorsperrow=2}

\author{Débora Souza, Patrícia Machado}

\affiliation{%
  \institution{Federal University of Campina Grande}
  \city{Campina Grande}
  \country{Brazil}
}

\renewcommand{\shortauthors}{Trovato et al.}

\begin{abstract}
LLM-based agents are becoming central to software engineering tasks, yet evaluating them remains fragmented and largely model-centric. Existing studies overlook how architectural components—such as planners, memory, and tool routers—shape agent behavior, limiting diagnostic power. We propose a lightweight, architecture-informed approach that links agent components to their observable behaviors and to the metrics capable of evaluating them. Our method clarifies what to measure and why, and we illustrate its application through real world agents, enabling more targeted, transparent, and actionable evaluation of LLM-based agents.
\end{abstract}



\received{20 February 2007}
\received[revised]{12 March 2009}
\received[accepted]{5 June 2009}

\maketitle

\section{Introduction}



Large Language Models (LLMs) have been explored in software engineering (SE) tasks, offering support for code writing, maintenance, documentation, and test automation~\cite{belzner2023large, jiang2024surveylargelanguagemodels}. When LLMs are endowed with the ability to interact with external tools, they transcend traditional text generation and function as agents capable of reasoning, decision-making, and performing complex, multi-step actions~\cite{wang2024survey, bass2025engineering}. This agent perspective enables LLMs to handle tasks that require contextual understanding, sequential decision-making, and multi-step execution in dynamic environments, opening new avenues for human–AI collaboration in software development.


As LLM-based agents — such as SWE-Agent~\cite{yang2024sweagent}, MetaGPT~\cite{hong2023metagpt}, and HuggingGPT~\cite{shen2023hugginggpt} — are transitioning from research prototypes to real-world applications, making structured and diagnostic evaluation a necessary prerequisite for reliability, safety, and effectiveness. Traditional software engineering methods, such as Test-Driven Development (TDD) and Behavior-Driven Development (BDD), are inadequate for these systems, as they assume deterministic behavior, fixed requirements, and binary pass/fail testing~\cite{xia2025edd}. LLM-based agents perform complex tasks involving reasoning, planning, tool use, memory management, and collaboration~\cite{durante2024agentaisurveyinghorizons}, while operating under evolving objectives, producing non-deterministic outputs, and exhibiting behaviors beyond fixed test cases~\cite{mohammadi2025agents}. This diversity illustrates the complexity and emerging capabilities of LLM-based agents, motivating careful consideration of how these systems are studied and assessed.


The work of Mohammadi et al.~\cite{mohammadi2025agents} propose a taxonomy distinguishing Evaluation Objectives—such as behavior, capabilities, reliability, and alignment—from the Evaluation Process, which includes interaction modes, data types, metric computation, and tooling. Similarly, Xia et al.~\cite{xia2025edd} introduce Evaluation-Driven Development (EDD), applying Test-Driven Development principles to support continuous, lifecycle-wide evaluation. Although other studies~\cite{hasan2025empiricaltestingpractices} have advanced the evaluation of LLM-based agents, the core problem remains the lack of a clear link between observable behaviors, agent architecture, and evaluation metrics.

LLM-based agents consist of interdependent architectural components -- such as planners, tool routers, and memory modules -- whose interactions give rise to observable behaviors (e.g., planning consistency, tool-use decisions, memory retrieval). These behaviors must be assessed through evaluation metrics rather than isolated component tests~\cite{liu2024agentdesignpatterncatalogue}. While EDD emphasizes considering architecture, there is no systematic method to map metrics to specific behaviors. Existing frameworks (e.g., DeepEval\footnote{\url{https://deepeval.com/}}, LangSmith\footnote{\url{https://docs.langchain.com/langsmith/observability}}) implement metrics but lack guidance on how to map them to specific behaviors within the agent architecture. As a result, current evaluations remain fragmented and ad hoc, often failing to diagnose which component failed and why.


In this work, we propose a systematic method for evaluating LLM-based agents that begins with the identification of observable behaviors, maps these behaviors to architectural components responsible for their emergence, and then defines metrics that meaningfully assess them. This approach provides a conceptual structure that clarifies the role of each component and demonstrates how its operation can be examined using existing evaluation tools. By linking observable behaviors to components and metrics, our method enables more organized and precise diagnosis of agent performance. This reduces misinterpretations that arise from purely system-level evaluation and supports a deeper understanding of how agents perceive, decide, act, and learn.

To this end, we (i) define what constitute architectural components and observable behaviors in LLM-based agents; (ii) propose a mapping that links behavioral phenomena, components, and evaluation metrics, grounded in established definitions; and (iii) demonstrate this approach on real-world agents, showing how to identify components, derive behaviors, and select appropriate metrics from existing frameworks.

\section{Related Work}

\textbf{Architectural components and patterns.}
LLM-based agents have evolved from standalone generative models into compound systems composed of out-of-model components such as interaction engineering, memory, planning and reflection, tool-use modules, and responsible AI mechanisms~\cite{mdpi2025agentic, wang2025aiagenticprogrammingsurvey, lu2024responsiblegenerativeaireference}. These works establish the architectural building blocks of agents, but they do not systematically relate components to the observable behaviors produced during execution.
\textbf{Evaluation objectives and processes.}
Evaluation practices remain dominated by aggregate end-task metrics. An analysis reports that over 66\% of studies rely almost exclusively on high-level accuracy or success-rate measures~\cite{xia2025edd}. This pattern holds across domains such as code generation (Pass@k, execution accuracy~\cite{jin2025llmsllmbasedagentssoftware, yehudai2025survey}) and requirements engineering (precision, recall, F1).
\textbf{Linking architecture and evaluation.}
More structured attempts—such as AgentArcEval, which organizes evaluation around quality attributes~\cite{qinghualuEval}, and Evaluation-Driven Adaptation, which links evaluation feedback to design refinement~\cite{xia2025edd}—highlight the need for methodological rigor. Yet modern tooling (e.g., LangSmith, DeepEval) provides granular metrics without guidance on how to map them to specific components or behaviors within the agent architecture. This leaves a methodological gap that our work targets by systematizing the explicit mapping on Component, Behavior and Metric.

\section{Research Questions}

The proposed research work aims to address the following research
questions.

\begin{enumerate}[label=\textbf{RQ\textsubscript{\arabic*}}]
\item[\textbf{RQ$_{1}$}] \textbf{How agent architectural components are related to observable behaviors?}
Rationale: The literature presents dozens of LLM-based agent architectures, but there is still no consolidated taxonomy that clarifies how components such as persona, memory, planning, tools, MCPs, etc contribute to agent behavior.
To answer this question, we will conduct a conceptual analysis and systematization of the literature (scientific + grey literature + codebases), aiming to define each component, its responsibilities, and the behaviors it generates.

\item[\textbf{RQ$_{2}$}] \textbf{How can observable behaviors and architectural components systematically mapped to existing evaluation metrics?}  
Rationale: Current evaluation practices are largely ad hoc and metric-driven rather than architecture-driven. A systematic mapping between (i) behavior, (ii) component and (iii) metric is missing.
We will develop a design-science-based mapping framework that associates behaviors with metric candidates drawn from tools such as LangSmith, DeepEval, and Opik. The goal is not to propose new metrics, but to clarify how existing metrics can meaningfully probe specific architectural functions.

\item[\textbf{RQ$_{3}$}] \textbf{To what extent does the proposed architecture-driven method improve structure, reproducibility, and diagnostic power in the evaluation of LLM-based agents?}

Rationale: To validate the method, we will apply it to real world publicly available agents such as SWE-Agent and MetaGPT. We will examine whether the method improves (i) diagnostic clarity regarding which component fails, (ii) reproducibility of evaluation procedures, and (iii) reduction of arbitrary metric choices compared to common evaluation practices.
This question will be answered through an experimental case study and comparative analysis.

\end{enumerate}

\section{Contribution}

This research provides two main contributions. First, we consolidate and formalize the architectural foundations of LLM-based agents by producing a unified taxonomy of components and the observable behaviors they generate (\textbf{RQ$_{1}$}). Unlike prior work that defines components in isolation, our contribution explicitly links behavioral signals—such as memory recall, planning steps, and tool/MCP usage—to the architectural elements that generate them. This taxonomy is grounded in scientific literature, grey literature, and code-level examination of real-world agents, offering the first architecture-to-behavior map for LLM-based agents.

Second, we introduce an architecture-informed evaluation method that systematically connects behaviors and components to existing metrics (\textbf{RQ$_{2}$}), enabling structured, reproducible, and diagnostic agent evaluation by making metric selection explicit and behavior-driven rather than ad hoc. The method operationalizes the component, behavior and metric mapping using metrics from LangSmith, DeepEval, Opik, and related tools, without proposing new metrics. We empirically validate this method by applying it to real-world agents such as SWE-Agent, demonstrating improved diagnostic clarity, clearer attribution of failures, and reduced arbitrariness in metric selection compared to common practice (\textbf{RQ$_{3}$}).

\section{Evaluation Plan}

The evaluation follows a comparative case-study design to assess whether the architecture-informed method improves structure, reproducibility, and diagnostic power (\textbf{RQ$_{3}$}). First, we consolidate the artifacts produced in \textbf{RQ$_{1}$} and \textbf{RQ$_{2}$}—namely, (i) the observable behaviors and its associated architectural taxonomy, and (ii) the behavior → component → metric mapping framework. These artifacts define the evaluation criteria.

Next, we instrument a representative set of open-source agents with process-level logging to capture intermediate behaviors such as planning traces, tool-use events, memory retrieval, and reflection steps. This instrumentation enables fine-grained behavioral observation. The core validation consists of a controlled comparison:
(1) Baseline (Ad Hoc Evaluation). The agent is evaluated using common high-level metrics (e.g., success rate, accuracy) selected without architectural considerations.
(2) Architecture-Informed Evaluation. The same agent is evaluated using metrics systematically selected to probe specific components (e.g., plan-consistency metrics for the Planner, tool-selection accuracy for the Tool Executor, consistency metrics for Memory/Persona).

Finally, diagnostic accuracy, clarity of failure attribution, and reproducibility of the evaluation process are compared between the two evaluation conditions described above. Improvements along these dimensions provide evidence that the architecture-informed method yields more structured and interpretable evaluations of LLM-based agents. This evaluation is subject to practical limitations. Agent architectures are heterogeneous, which may affect the generality of the method. The analysis also depends on the availability and quality of logging and tooling, which are not uniformly supported across agents.

\bibliographystyle{ACM-Reference-Format}

\appendix

\end{document}